\documentclass[%
 reprint,
superscriptaddress,
 amsmath,amssymb,
 prl,
floatfix,
]{revtex4-1}
\pdfoutput=1

\usepackage{graphicx}
\usepackage{dcolumn}
\usepackage{bm}
\usepackage[colorlinks=true,
            linkcolor=blue,
            urlcolor=blue,
            citecolor=blue]{hyperref}
\usepackage[mathlines]{lineno}
\usepackage[all]{hypcap}
\usepackage{SIunits}
\usepackage{upgreek}
\usepackage[artemisia]{textgreek}
\usepackage{tabularx}

\begin{document}

\preprint{APS/123-QED}

\title{Pump-Depletion Dynamics and Saturation of Stimulated Brillouin Scattering in Shock Ignition Relevant Experiments}

\author{S. Zhang}
\affiliation{Department of Mechanical and Aerospace Engineering, University of California San Diego, La Jolla, California 92093, USA}

\author{J. Li}
\affiliation{Department of Mechanical and Aerospace Engineering, University of California San Diego, La Jolla, California 92093, USA}
 
\author{C. M. Krauland}
\affiliation{Inertial Fusion Technology, General Atomics, San Diego, California 92121, USA}

\author{F. N. Beg}
\affiliation{Department of Mechanical and Aerospace Engineering, University of California San Diego, La Jolla, California 92093, USA}

\author{S. Muller}
\affiliation{Inertial Fusion Technology, General Atomics, San Diego, California 92121, USA}

\author{W.~Theobald}
\affiliation{Laboratory for Laser Energetics, University of Rochester, Rochester, New York 14623, USA}

\author{J.~Palastro}
\affiliation{Laboratory for Laser Energetics, University of Rochester, Rochester, New York 14623, USA}

\author{T.~Filkins}
\affiliation{Laboratory for Laser Energetics, University of Rochester, Rochester, New York 14623, USA}

\author{D. Turnbull}
\affiliation{Laboratory for Laser Energetics, University of Rochester, Rochester, New York 14623, USA}

\author{D. Haberberger}
\affiliation{Laboratory for Laser Energetics, University of Rochester, Rochester, New York 14623, USA}

\author{C.~Ren}
\affiliation{Laboratory for Laser Energetics, University of Rochester, Rochester, New York 14623, USA}
\affiliation{Department of Mechanical Engineering, University of Rochester, Rochester, New York 14623, USA}

\author{R.~Betti}
\affiliation{Laboratory for Laser Energetics, University of Rochester, Rochester, New York 14623, USA}
\affiliation{Department of Mechanical Engineering, University of Rochester, Rochester, New York 14623, USA}

\author{C.~Stoeckl}
\affiliation{Laboratory for Laser Energetics, University of Rochester, Rochester, New York 14623, USA}

\author{E. M. Campbell}
\affiliation{Laboratory for Laser Energetics, University of Rochester, Rochester, New York 14623, USA}

\author{J. Trela}
\affiliation{Université de Bordeaux-CNRS-CEA, CELIA (Centre Lasers Intenses et Applications) UMR 5107 F-33400 Talence, France}

\author{D. Batani}
\affiliation{Université de Bordeaux-CNRS-CEA, CELIA (Centre Lasers Intenses et Applications) UMR 5107 F-33400 Talence, France}

\author{R. Scott}
\affiliation{Rutherford Appleton Laboratory, Science and Technology Facilities Council, UK}

\author{M.~S.~Wei}%
\email{mingsheng.wei@rochester.edu}
\affiliation{Inertial Fusion Technology, General Atomics, San Diego, California 92121, USA}
\affiliation{Laboratory for Laser Energetics, University of Rochester, Rochester, New York 14623, USA}

\date{\today}

\begin{abstract}
As an alternative inertial confinement fusion scheme with predicted high energy gain and more robust designs, shock ignition requires a strong converging shock driven by a shaped pulse with a high-intensity spike at the end to ignite a pre-compressed fusion capsule. Understanding nonlinear laser-plasma instabilities in shock ignition conditions is crucial to assess and improve the laser-shock energy coupling. Recent experiments conducted on the OMEGA-EP laser facility have for the first time demonstrated that such instabilities can $\sim$100\% deplete the first 0.5 ns of the high-intensity laser pump. Analysis of the observed laser-generated blast wave suggests that this pump-depletion starts at 0.01--0.02 critical density and progresses to 0.1--0.2 critical density. This pump-depletion is also confirmed by the time-resolved stimulated Raman backscattering spectra. The dynamics of the pump-depletion can be explained by the breaking of ion-acoustic waves in stimulated Brillouin scattering. Such strong pump-depletion would inhibit the collisional laser energy absorption but may benefit the generation of hot electrons with moderate temperatures for electron shock ignition [Shang et al. Phys. Rev. Lett. 119 195001 (2017)]. 

\end{abstract}

\maketitle


Shock ignition (SI) is an alternative inertial confinement fusion (ICF) scheme~\cite{betti2007shock,betti2008}. SI utilizes an initial nanoseconds pulse at low intensities ($10^{14}\hbox{--}10^{15}~{\rm W/cm^2}$) to irradiate and compress a cryogenic DT capsule. Subsequently, a higher intensity spike ($\sim$10$^{16}~{\rm W/cm^2}$, $\sim$0.5~ns) generates a spherical converging shock to form a non-isobaric hot spot and ignite the compressed fuel~\cite{NIFSI2009,Anderson2013NIF,LMJSI,Ribeyre2009PPCF,SIHiPER}. The advantage of this scheme is that the separation of compression and ignition phases may provide a more stable implosion and higher energy gain than the conventional central hot spot ICF ignition~\cite{AtzeniReview2014}. However, the laser-shock energy coupling remains uncertain, since the spike pulse may lose energy due to laser-plasma instabilities (LPIs)~\cite{kruer2003physics}, such as stimulated Raman scattering (SRS), stimulated Brillouin scattering (SBS)~\cite{SRSLiu}, and two-plasmon decay (TPD)~\cite{TPDSimon}. These instabilities can convert laser energy into electron plasma waves, ion-acoustic waves (IAW) and scattered light. Over the entire duration of the SI high-intensity spike pulse, all the LPIs are nonlinear, and yet to be explored experimentally~\cite{Batani2014review}. 

The large-scale coronal plasma created by the implosion pulse can significantly impact the role of each LPI mode~\cite{Rosenberg2018}. SI with megajoule lasers would have the coronal plasma with a scale length $L_n\sim 300\hbox{--} 500$~\micro\meter{} and electron temperature $T_{\rm e}> 3$~keV. Particle-in-cell (PIC) simulations with SI high intensity and large plasmas have shown $>$50\% SBS reflectivity~\cite{Klimo2011,Klimo2013,Hao2016}, which is not seen in small-scale simulations~\cite{Weber2015,Riconda2011,Klimo20142D} or in experiments~\cite{theobald2015spherical,theobald2012spherical,Depierreux2009,Depierreux2012,Baton2012,Baton2017,Goyon2013,Cristoforetti2018}. Some experiments have observed a burst of SBS at the beginning of the laser spike~\cite{theobald2015spherical,Depierreux2009,Depierreux2012,Baton2017}. However, those experiments were limited by either small plasma scale-lengths ($L_n<170$~\micro\meter{}) or low laser intensity ($\sim10^{14}\hbox{--}10^{15}~{\rm W/cm^2}$). It is also challenging to extend the PIC simulations to full time and spatial scale due to computational limitation. Therefore, experiments are warranted to characterize the LPIs in the interaction between a $10^{16}~{\rm W/cm^2}$ laser and a large-scale plasma.

This Letter reports on a series of experiments to study the laser propagation and LPI physics with SI-relevant high intensity ($\sim$10$^{16}~{\rm W/cm^2}$) in large-scale ($L_n\sim260\hbox{--}330$~\micro\meter{}) keV plasmas, a previously unexplored regime. We observed unprecedentedly strong pump-depletion that started at the low-density ($n_{\rm e}\sim 0.01 n_{\rm c}$) plasma and progressed into the higher density ($n_{\rm e}>0.1 n_{\rm c}$) region. Such pump-depletion dynamics observed for the first time can be explained by the local SBS saturation induced by IAW-breaking. This SBS saturation mechanism help resolve the long-standing discrepancy on the reflectivity data between PIC simulations~\cite{Klimo2011,Klimo2013,Hao2016} and previous experiments~\cite{theobald2015spherical,theobald2012spherical,Depierreux2009,Depierreux2012,Baton2012,Baton2017,Cristoforetti2018}. These new findings have significant implications for SI. 

\begin{figure}[t]
\includegraphics{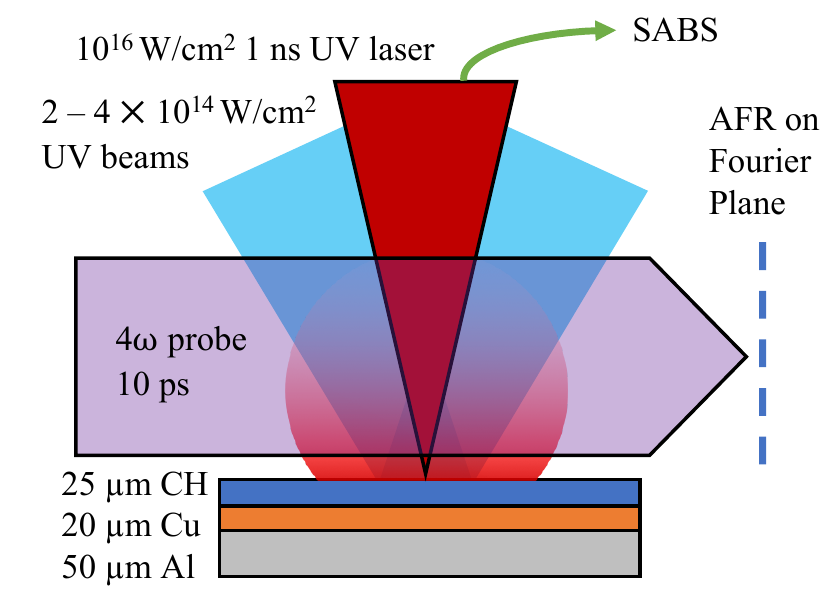}
\caption{\label{fig:setup} The experimental setup. One-two 2 kJ UV laser beams (blue) irradiated the disk target to generate the plasma. One tight focused 1-ns 1.25~kJ UV laser (red) was injected to interact with the plasma. A Sub-Aperture Backscattering Spectrometer (SABS) was used to diagnose the SRS backscattering driven by the interaction beam.  A 4$\omega$ laser probed the plasma immediately after the laser plasma interaction. The filter rings (blue dashes) of the Angular Filter Refractometer (AFR) were placed at the Fourier plane of the probe beam to image the refraction map.  }
\end{figure}

The experiments were performed on the OMEGA-EP laser facility~\cite{OMEGAEP} at the Laboratory for Laser Energetics, Rochester, US. To produce a shock ignition relevant plasma, one or two 2~kJ, 2~ns UV lasers irradiated a 3-layer disk target (25~\micro\meter{} CH/20~\micro\meter{} Cu/50~\micro\meter{} Al), as shown in Fig.~\ref{fig:setup}. These 2-ns lasers were smoothed by 750~\micro\meter{} 8$^{\rm th}$-order super-Gaussian distributed phase plates. Delayed relative to the start of the long pulse beams by 1.0--1.5~ns, a $10^{16}~$W/cm$^2$ (1~ns, 1.25~kJ) UV interaction pulse was then injected into the plasma along the target normal direction. The beam was tightly focused on the target surface with a $\sim$80~\micro\meter{} diameter spot without a distributed phase plate. It created a conical blast wave expanding radially from the laser axis when propagating in the large-scale coronal plasma. Immediately after the interaction pulse, a 10 ps 4$\omega$ ($\lambda = 263$~nm) laser probed the plasma to image the blast wave on a refraction map using Angular Filter Refractometer (AFR)~\cite{AFR}. We also used the streaked Sub-Aperture Backscattering Spectrometer (SABS) to diagnose the temporally resolved spectrum of the SRS backscattered light (430--750~nm). The radiation-hydrodynamic code FLASH~\cite{FLASH2000,Dubey2009512} was used to simulate the plasma profiles (electron temperature $T_{\rm e}$ and electron density $n_{\rm e}$). The simulations were benchmarked in previous experiments~\cite{ShuPoP}.

\begin{figure}[t]
\includegraphics{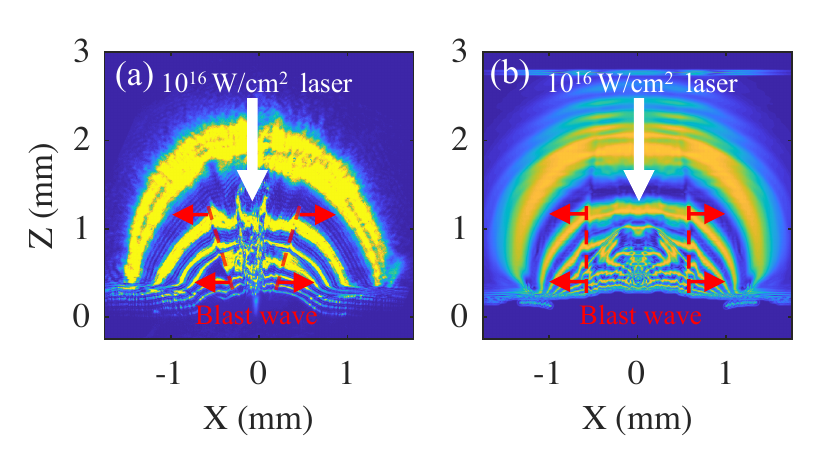}
\caption{\label{fig:AFR} (a) The experimental AFR image; (b) the FLASH simulated AFR image. The target surface is at $z = 0$. The red dashed lines mark the blast waves. White arrows show the directions of the high intensity UV interaction laser.  Red arrows represent the moving directions of the blast waves. The difference of the blast wave shapes (conical vs. cylindrical) can be caused by the LPI-induced pump-depletion (see discussions in the text).}
\end{figure}

The shape of the blast waves, shown as the red dashed lines in Fig.~\ref{fig:AFR}, indicate that the high intensity laser is strongly pump-depleted. Fig.~\ref{fig:AFR}(a) is the experimental AFR image captured at the end of the interaction laser. In this image, the interaction laser was delayed by 1.5 ns relative to the start of the 2 kJ initial long pulse beam. At the start of the interaction pulse, the plasma has $T_{\rm e}\sim1.5$~keV and exponential density length scale $L_n\sim330$~\micro\meter{} in the region between $n_{\rm c}/10$ and $n_{\rm c}/4$. In Fig.~\ref{fig:AFR}(a), the blast wave front has a conical shape. The diameter of the cone at the laser entrance ($z=1.2$~mm, $n_{\rm e} \sim0.02 n_{\rm c}$) is 90\% larger than the diameter at $z=0.4$~mm ($n_{\rm e}\sim0.1 n_{\rm c}$). We compared the experimental image with FLASH simulations to understand the pump-depletion effects. The simulations have the same $f/6.5$ focusing laser corresponding to the experiments but did not include any LPI physics.  The synthetic AFR image is shown in Fig.~\ref{fig:AFR}(b). Instead of a conical shape, the simulation shows a cylindrical blast wave. At $z=1.2$~mm, the experimental blast wave has a similar diameter to the simulation (0.96 mm versus 1.14 mm), but at $z = 0.4$~mm, the experimental diameter is only 42\% (0.50~mm versus 1.19~mm) of the one shown in the simulation. Based on Sedov's self-similar blast wave model~\cite{Sedov}, the radius of the cylindrical blast wave can be described as $r\propto t^{1/2} (E/\rho)^{1/4}$, where $t$ is the time after the explosion, $E$ is the absorbed laser energy per axial length and $\rho$ is the medium density. This model suggests that the smaller blast wave radius at the deep plasma region can be due to reduced laser energy, higher density, or a shortened laser interaction time. To elucidate the underlying cause, FLASH simulations in 1-D cylindrical geometry were performed to examine the blast wave dependence on those parameters. We found that only a shortened laser interaction time can explain this conical blast wave with the reduced diameter in the high density plasma region.

\begin{figure}[t]
\includegraphics{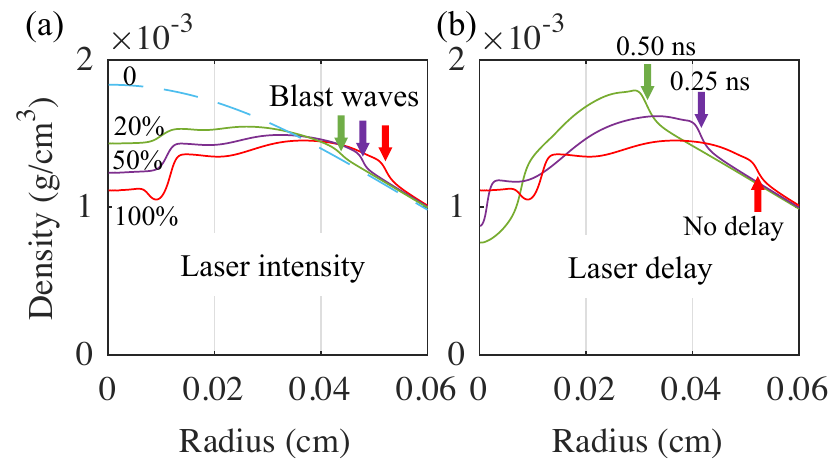}
\caption{\label{fig:1D} The simulated density profiles after the laser interaction (a) when varying the intensity, and (b) with different laser delays. The arrows mark the fronts of the blast wave.}
\end{figure}

In the 1D blast wave simulations, we scanned the laser intensity and the laser duration respectively.  The initial plasma profiles are extracted from the lineout in the radial direction at $z = 0.5$~mm of the 2D FLASH simulation. The density profile after the laser interaction is shown in Fig.~\ref{fig:1D}(a) (for varying intensity) and Fig.~\ref{fig:1D}(b) (for varying pump-depletion time). As shown in Fig.~\ref{fig:1D}(a), the laser intensity weakly affects the position of the blast wave. When the intensity is reduced from 100\% to 20\% of the original, the blast wave radius decreases by 18\% from $520\pm10$~\micro\meter{} (red line) to $430\pm 10$~\micro\meter{} (green line), which is still 50\% larger than the experimental radius at $z = 0.5$~mm ($r=290\pm30$~\micro\meter{}).  The density gradient at the wave front also decreases with the intensity. When the intensity is further reduced to 5\% of the original (profile not shown), the wave front becomes unobservable. No simulation with reduced intensity can reproduce the small but clear blast wave found in the experiment. On the other hand, shorter laser pulse duration can effectively reduce the driven blast wave radius while keeping the wave front as observable. As shown in Fig.~\ref{fig:1D}(b), when the first 0.25~ns (purple line) or 0.50~ns (green) of the laser is fully pump-depleted---the remaining 0.75~ns or 0.50~ns pulse has the original intensity---the radius $r$ decreases linearly from $520\pm10$~\micro\meter{} ($r_0$) to $420\pm10$~\micro\meter{} or $320\pm10$~\micro\meter{}, respectively. The relative change of the blast wave radius $(r_0-r)/r_0$  is proportional to the pump-depletion time ($t_{\rm PD}$) as,
\begin{equation}\label{eq:pdtime}
   \frac{r_0-r}{r_0}=\frac{t_{\rm PD}}{1.3~{\rm ns}}.
\end{equation}
In conclusion, the LPI-induced pump-depletion blocks the first part of the laser. After the pump-depletion, the interaction laser continues to propagate with the same order of the original intensity and generates a small blast wave. 

\begin{figure}[t]
\includegraphics{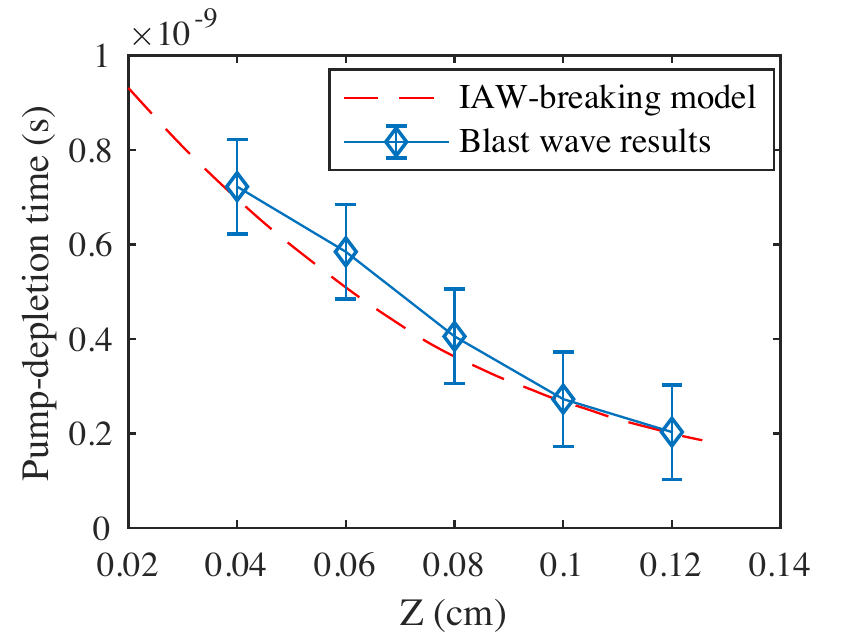}
\caption{\label{fig:time} The pump-depletion times ($t_{\rm PD}$'s) at different distances above the target (Z). Blue diamonds are the experimental $t_{\rm PD}$'s from the blast wave analyses. Dashed red lines show the $t_{\rm PD}$'s predicted by the IAW breaking model. }
\end{figure}

With the linear blast wave model Eq.~(\ref{eq:pdtime}) shown above, we can calculate the pump-depletion time $t_{\rm PD}$ along the laser axis using the ratio of experimental radius to simulated radius   $r_{\rm exp}(z)/r_{\rm sim}(z)$ as,
\begin{equation}
   t_{\rm PD}(z)=1.3\bigg(1-\frac{r_{\rm exp} (z)}{r_{\rm sim}(z)}\bigg)~{\rm ns}.
\end{equation}
The $t_{\rm PD}$'s along $z$ are shown as the diamonds in Fig.~\ref{fig:time}. The pump-depletion time increases by 0.5~ns from $z = 1.2$~mm to $z = 0.4$~mm. The local pump-depletion time agrees with an IAW-breaking SBS saturation model shown as the red dashed line, which is discussed later in this paper.

\begin{figure}[b]
\includegraphics{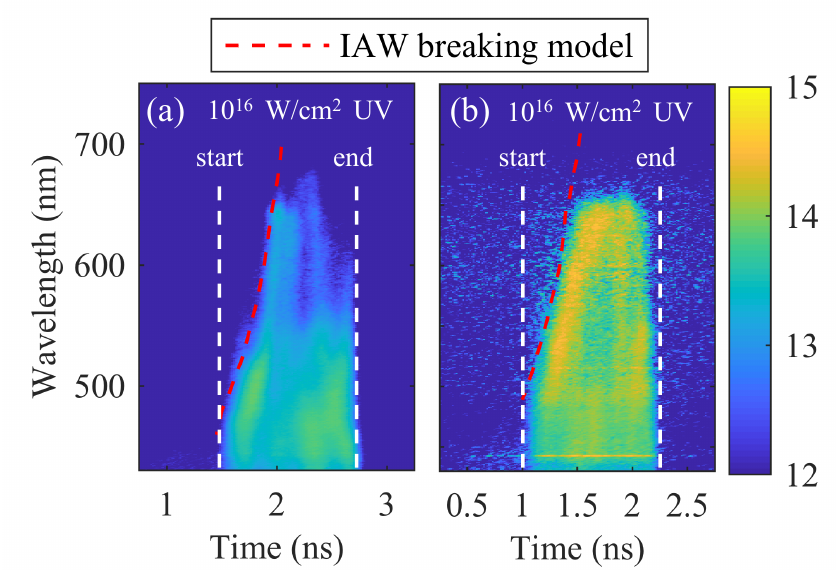}
\caption{\label{fig:SRS} Measured SRS spectra from two experiments in different plasma conditions. (a) The laser and plasma conditions are similar to the blast wave shot ($L_n\sim330$~\micro\meter{}, $T_{\rm e}\sim1.5$~keV); (b) The plasma has higher temperature (2.0~keV) and shorter scale length (260~\micro\meter{}), and the interaction beam is delayed by 1.0~ns. The color bar is in $\log_{10}$ scale. Dashed red lines are pump-depletion positions predicted by the IAW breaking model. 0 of the time axis is the start of the low-intensity UV lasers. }
\end{figure}

Furthermore, the time-resolved SRS spectra also show the delayed backscattering light from the deep region. Figure~\ref{fig:SRS} shows two SRS spectra from experiments in different plasma conditions. The image in Fig.~\ref{fig:SRS}(a) is under the same conditions as the blast wave image shown above ($L_n\sim 330$~\micro\meter{}, $T_{\rm e}\sim 1.5$~keV). The SRS signal from $n_{\rm e} < 0.02~n_{\rm c}$ low-density region ($\lambda\sim450$~nm) starts with the high-intensity UV laser. Then the SRS signal from the deeper region (0.02--0.20~$n_{\rm c}$) gradually shortens. The shot in Fig.~\ref{fig:SRS}(b) has a plasma with smaller $L_n \sim 260$~\micro\meter{} and higher $T_{\rm e}\sim2.0$~keV. The interaction laser is delayed by 1.0~ns relative to the start of the low intensity beams. Similar to the previous shot, the backscattered SRS light starts at the short wavelength ranging from 430~nm to 550~nm corresponding to $n_{\rm e}<0.05~n_{\rm c}$. 
For longer wavelengths (550--650 nm) scattered from higher densities (0.05--0.20 $n_{\rm c}$), the onset of SRS is delayed by up to 0.4 ns. 
These delays of the SRS signal from the deep region indicate that the first 0.4--0.5~ns of the laser pulse is fully depleted at the low-density region, which agrees with the delayed laser penetration as described in the blast wave analysis. The PIC simulations corresponding to these experiments suggest that strong SBS can grow in the $n_{\rm e}<0.05~n_{\rm c}$ region and suppress SRS and TPD in the higher density region~\cite{LiPRL}


 The inward movement of the pump-depletion front can be explained by the IAW-breaking in SBS. An IAW would break when the ion quiver velocity ($v_{\rm quiver}$) is close to the wave's phase velocity ($c_{\rm s}$), as $v_{\rm quiver}\approx c_{\rm s}$~\cite{Judice1973,Cohen1997}. There are two IAW modes in a CH plasma: a fast mode and a slow mode. The fast mode is dominated by H$^+$ ions, while C$^{6+}$ ions dominate the slow mode. The phase velocities of these two modes ($c_{\rm fast}$ and $c_{\rm slow}$) are calculated from the kinetic model, expressed as Eq.~(20) in Ref.~\onlinecite{Williams1995}. Under the conditions of our experiments, $c_{\rm fast}\sim1.1 \sqrt{T_{\rm e}/m_{\rm H}}$ and $c_{\rm slow}\sim(0.6\hbox{--}0.8)\sqrt{T_{\rm e}/m_{\rm H}}$. $v_{\rm quiver}$ can be calculated from the energy of the IAW ($E_{\rm IAW}$) since half of $E_{\rm IAW}$ is the ion kinetic energy and the other half is the potential energy. In strong SBS, in which the laser is fully scattered, IAW would constantly gain energy based on its wave frequency as,
\begin{equation}\label{eq:dEdT}
    \frac{{\rm d}E_{\rm IAW}}{{\rm d}t}=\frac{\omega_{\rm cs}}{\omega_0}P_{\rm L},
\end{equation}
where $\omega_{\rm cs}$ and $\omega_0$ are the angular frequencies of IAW and the laser, respectively, and $P_{\rm L}$ is the laser power. When the light is backscattered, the IAW's wave number $k_{\rm IAW}$ would approximately equal to $2\,k_{\rm L}$  to match the momentum conservation~\cite{kruer2003physics}, so
\begin{equation}\label{eq:wcsw0}
    \frac{\omega_{\rm cs}}{\omega_0}\approx\frac{2c_{\rm s}}{c}
\end{equation}
According to Eq.~(\ref{eq:dEdT}) and~(\ref{eq:wcsw0}), the IAW breaking condition of each IAW mode in a small volume $S_{\rm L} \Delta z$ can be reformed as the following: for the fast mode,
\begin{equation}\label{eq:tfastmode}
    \frac{c_{\rm fast}}{c} P_{\rm L} \Delta t\approx\frac{1}{2} N_{\rm H}m_{\rm H}c_{\rm fast}^2,
\end{equation}
and for the slow mode,
\begin{equation}\label{eq:tslowmode}
    \frac{c_{\rm slow}}{c} P_{\rm L} \Delta t\approx\frac{1}{2} N_{\rm C}m_{\rm C}c_{\rm slow}^2,
\end{equation}
where $N_{\rm H}$ and $N_{\rm C}$ are the numbers of H$^+$ and C$^{6+}$ ions in volume $S_{\rm L}\Delta z$, so $N_{\rm H}=N_{\rm C}=n_{\rm e} S_{\rm L}\Delta z/7$.
Here a square laser pulse with a constant power $P_{\rm L}$ is assumed,  and $\Delta t$ is each SBS's growing time in this volume and $S_{\rm L}$ is the laser cross-section at position $z$. When both modes grow simultaneously following Eq.~(\ref{eq:dEdT}), they would share the laser power $P_{\rm L}$. However, the slow mode would saturate about 8 times slower than the fast mode because of the large mass of the C$^{6+}$ ion,
so we only considered the slow mode when calculating the saturation of SBS. After all C$^{6+}$ ions in this volume are accelerated to the IAW phase velocity, SBS would stop amplifying the IAW and stop reflecting, so the laser can propagate into the next region. The pump-depletion front would move inward with speed $v_{\rm PD}$ expressed as
\begin{equation}\label{eq:vpd}
    v_{\rm PD}(z)=\frac{\Delta z }{\Delta t}\approx\frac{14P_{\rm L}}{c_{\rm slow}(z) n_{\rm e}(z)S_{\rm L}(z)\,c\,m_{\rm C}}.
\end{equation}
We use this model to predict the positions of the pump-depletion region, shown as the red dashed lines in Fig.~\ref{fig:time},~\ref{fig:SRS}(a) and~\ref{fig:SRS}(b). The three experiments have two different plasma conditions as described above. The calculations of $c_{\rm slow}(z)$ used $T_{\rm e}(z)$, $T_{\rm i}(z)$  and $n_{\rm e}(z)$ from the corresponding FLASH simulations. The laser cross section, $S_{\rm L}(z)=\pi(z/6.5+80~\micro\meter{})^2/4$, is the same in all calculations. This cross section represents an $f/6.5$ focusing laser with an 80~\micro\meter{} focal spot at $z=0$, which is consistent with the width of the central channel as shown in Fig.~\ref{fig:AFR}(a). We can see that, although the plasma conditions and the measurement methods are different, the predicted pump-depletion positions in the $0.02\hbox{--}0.20~n_{\rm c}$ region agree well with all three experimental results.

This pump-depletion dynamics can explain why most PIC simulations show stronger SBS than the experiments do. This discrepancy can be caused by the short time scale in PIC simulations and the small plasma length scale in previous experiments. The simulations in Ref.~\onlinecite{Klimo2011,Klimo2013,Hao2016} have large plasmas but the simulation times are not long enough for the SBS to saturate. Ref.~\onlinecite{Klimo2011} simulated a mm sized $\sim$2~keV CD plasma similar to the conditions of our experiments. Based on the pump-depletion moving speed $v_{\rm PD}(z)$ in Eq.~(\ref{eq:vpd}), the SBS would saturate after 500~ps, which is much longer than the simulation time (90~ps). Although Ref.~\onlinecite{Riconda2011} simulated a smaller (160~\micro\meter{}) D$^+$ plasma, the simulation time (5~ps) is still one order shorter than the calculated SBS saturation time from Eq.~(\ref{eq:tfastmode}). These simulations are still in the SBS growth phase, so the simulated high reflectivity is reasonable.  Compared to the experiments, the simulations would over predict the SBS reflectivity as the simulation times were much shorter than the SBS saturation time.  On the other hand, the small plasma scale and the low temperature in the previous experiments~\cite{theobald2015spherical,theobald2012spherical,Depierreux2009,Depierreux2012,Baton2012,Baton2017,Cristoforetti2018} may have been the cause of the low SBS reflectivity. Our experiments show that the pump-depletion in the $L_n \sim 140$~\micro\meter{} plasma is not as observable as that in the $L_n \geqslant 260$~\micro\meter{} plasma. In the small-scale experiment, the signals of TPD and SRS from $n_{\rm e}> 0.2~n_{\rm c}$ region appeared simultaneously with the interaction laser. Eq.~(\ref{eq:tfastmode}) and (\ref{eq:tslowmode}) also suggest that the SBS would saturate faster in a low-temperature small-scale plasma, thus, lower SBS reflectivity is expected in previous experiments. 

Although large-scale PIC simulations have not shown the SBS saturation, small-scale or hybrid PIC simulations have shown that the high SBS reflectivity drops to a few percent after a short period~\cite{Weber2005PRL,Weber2005PoP,DivolHybridPIC,Kruer1975}. Ref.~\onlinecite{Weber2005PRL,Weber2005PoP} have presented a PIC simulation of a $10^{16}~{\rm W/cm^2}$ 1-\micro\meter{} laser interacting with a 40~\micro\meter{} thick 0.3$n_{\rm c}$ H$^+$ plasma. The reflectivity dropped 6.5~ps after the start of the laser. The simulation also showed IAW-breaking. Based on Eq.~(\ref{eq:tfastmode}), our model predicts that the IAWs would break at 6.2~ps in this condition, which agrees well with the reflectivity dropping time in the simulation.

This strong pump-depletion can affect the laser-shock energy coupling in multiple ways. First, the pump-depletion can block the laser from reaching the high density (up to $n_{\rm c}$) region, where the collisional absorption dominates. As a result, the collisional absorption would be greatly reduced. On the other hand, it is possible to exploit LPI-induced hot electrons to generate the strong shock for SI~\cite{gus2012ablation,aisa2015dense,aisa2016,Shang2017}. As shown in our experiment, the strong pump-depletion can block the laser from reaching the 0.25~$n_{\rm c}$ region, where TPD can generate $>$100~keV hot electrons. SRS in low density ($<$0.2~$n_{\rm c}$) region only generate hot electrons with low energies. As the second effect, the pump-depletion can lower the hot electron temperature. This may explain why the hot electron temperatures are lower in our experiments~\cite{ShuMJ} ($T_{\rm hot}\sim 40$~keV) than those in small-scale experiments~\cite{theobald2015spherical,nora2015gigabar} (60--70~keV). This effect favors the electron shock ignition since the low-$T_{\rm hot}$ electrons were predicted to be able to generate the ignition shock more efficiently~\cite{aisa2015dense}.

In conclusion, the first experiments to characterize LPI at full-scale shock ignition-relevant laser intensity and plasma conditions have shown evidence of strong pump-depletion of the spike pulse, which is in contrast with previous smaller-scale experiments where SBS was suppressed. This pump-depletion was observed to start at the $0.01\hbox{--}0.02~n_{\rm c}$ low-density region and progress into $0.1\hbox{--}0.2~n_{\rm c}$ region over the first 0.5~ns of the spike pulse. This dynamic agrees with the IAW-breaking SBS saturation model. This SBS saturation mechanism can explain the reflectivity discrepancy between previous PIC simulations and experiments, where simulations with significantly shorter time-scale overpredicted SBS and low SBS is expected in previous experiments with either low intensities or small-scale plasmas. The IAW-breaking may further perturb the plasma and impact LPI~\cite{Li2017}, which has not been considered in SI-scheme so far. Furthermore, the strong pump-depletion would inhibit the collisional laser absorption in the megajoule-scale SI scheme, but may benefit electron shock ignition by reducing the TPD generated high energy electrons. Effects of the overlapped beams on LPI and hot electron generation in shock ignition require further investigation. 

We acknowledge the OMEGA-EP laser facility staff at the Laboratory for Laser Energetics. This work was performed under the auspices of U.S. DOE NNSA under the NLUF program with award number DE-NA0002730, DE-NA0003600, DOE Office of Science under the HEDLP program with award number DE-SC0014666 and DOE Office of Science under DE-SC0012316. The FLASH code used in this work was in part developed by the DOE NNSA-ASC OASCR Flash Center at the University of Chicago. S.Z. thanks Maylis Dozieres, Adam Higginson, and Krish Bhutwala for editing the manuscript. The support of the DOE does not constitute an endorsement by the DOE of the views expressed in this article.


%

\end{document}